# Wavelength-Multiplexed Entanglement Distribution over 10 km of Fiber

Han Chuen Lim (1,2), Akio Yoshizawa (2,3), Hidemi Tsuchida (2,3), Kazuro Kikuchi (1)
1 : Graduate School of Engineering, The University of Tokyo, 7-3-1 Hongo, Bunkyo-ku, Tokyo, 113-8656, Japan.
Email: hanchuen@ginjo.t.u-tokyo.ac.jp
2 : Photonics Research Institute, National Institute of Advanced Industrial Science and Technology (AIST). 1-1-1, Umezono, Tsukuba, Ibaraki, 305-8568, Japan.
3 : CREST, Japan Science and Technology Agency (JST). 4-1-8, Honcho, Kawaguchi, Saitama, 332-0012, Japan.

**Abstract**

*We report the first experimental demonstration of wavelength-multiplexed entanglement distribution. 44 channels of highly-entangled photon-pairs from one single broadband source are distributed over 10 km of fiber.*

**Introduction**

In future quantum communication applications such as multi-party quantum cryptography [1], quantum secret sharing [2] and distributed quantum computing [3], application users are required to share and consume quantum entanglement as a resource. However, the total amount of shared entanglement cannot be increased via local operations and classical communications (LOCC) [4], and so sharing of entanglement among distantly located users must definitely involve some form of entanglement distribution.

Recently, we have presented the concept of a *local-area entanglement distribution fiber network* [5], in which a centrally located service provider produces highly-entangled photon-pairs via spontaneous parametric down-conversion (SPDC), and distributes these photon-pairs over fiber-optic transmission lines to application users. Experimental progress along this direction includes realizations of high quality telecom-band entangled photon-pair sources [6-10], and demonstrations of entanglement distribution over 100 km of optical fiber [11-14]. However, most of the entangled photon-pair sources demonstrated to date have exhibited a relatively narrow bandwidth as compared to the available transmission bandwidth. To fully utilize the transmission bandwidth of optical fiber, a service provider would have to wavelength-multiplex many such narrowband sources before transmission, and this is not cost-effective.

In another recent paper [15], we have proposed and demonstrated experimentally a *single* broadband source that is well-suited for multi-channel wavelength-multiplexed entanglement distribution. The idea is to employ a 1-mm-long periodically-poled lithium niobate (PPLN) waveguide operating near degeneracy wavelength for SPDC. A simple calculation reveals that the SPDC bandwidth of such a short waveguide could cover hundreds of nm and therefore it is a promising candidate as an ultra-broadband source of entangled photon-pairs. In this work, we demonstrate wavelength-multiplexed entanglement distribution for the first time using the proposed source. Forty-four wavelength channels of highly-entangled photon-pairs are produced from a single source and distributed over 10 km of fiber.

**Experiment**

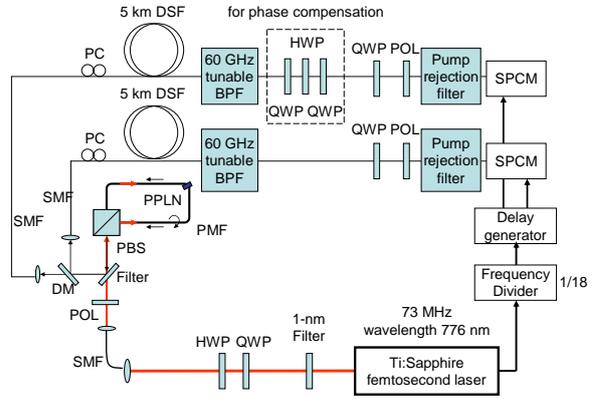

*Figure 1: Experimental setup. BPF: band-pass filter, DM: dichroic mirror, DSF: dispersion-shifted fiber, HWP: half-wave plate, PBS: polarization beam-splitter, PC: polarization-controller, PMF: polarization-maintaining fiber, POL: polarizer, PPLN: periodically-poled lithium niobate waveguide, QWP: quarter-wave plate, SMF: single-mode fiber, SPCM: single-photon counter module.*

Figure 1 shows the experimental setup. The source consists of a 1-mm-long MgO-doped type-0 quasi-phase-matched PPLN waveguide (HC Photonics) placed at the center of a polarization-diversity loop. For a description on the principle of this source, see [10]. The pump laser is a Ti:Sapphire femtosecond laser operating at 776 nm. A 1-nm-bandwidth filter reduces the pump spectral width to about 1 nm. This improves the quality of the output entangled photon-pairs significantly [15]. The photon-pairs produced are entangled in polarization and can be described by $|H\rangle_s|H\rangle_i + e^{i\theta}|V\rangle_s|V\rangle_i$, where $H$ and $V$ denote horizontal and vertical polarizations, respectively. The subscripts $s$ and $i$ denote signal and idler, respectively, and $\theta$ is an unknown but constant phase.

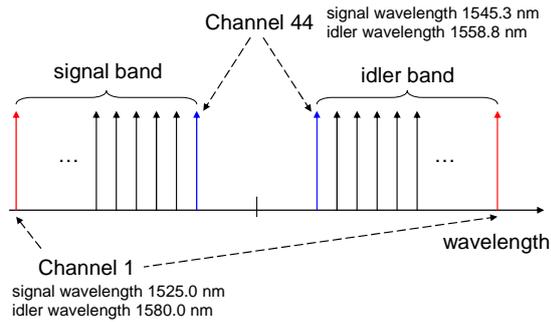

*Figure 2: Each channel consists of a signal channel and an idler channel. Channel 1's signal wavelength is 1525.0 nm and channel spacing is 60 GHz.*

A dichroic mirror (DM) separates signal and idler photons into two bands, as shown in Fig. 2. Both bands are then sent over 5-km-long dispersion-shifted fibers (DSFs). For wavelength-demultiplexing, we have used a pair of band-pass filters (BPFs, Optoquest) having 60 GHz passband and tunable from 1520 to 1580 nm. In a practical system, arrayed waveguide gratings (AWGs) should be used. Density matrices of the demultiplexed photon-pairs are obtained by quantum state tomography [16]. More details of this experiment can be found in [10].

Although our source can operate without temperature control, we have used a thermo-electric controller to set the waveguide temperature to 20.0 degree Celsius for enhanced stability in this experiment. Two quarter-wave plates (QWPs) and a half-wave plate (HWP) are placed immediately after the BPF of the signal channel for compensating the unknown phase.

**Results and Discussion**

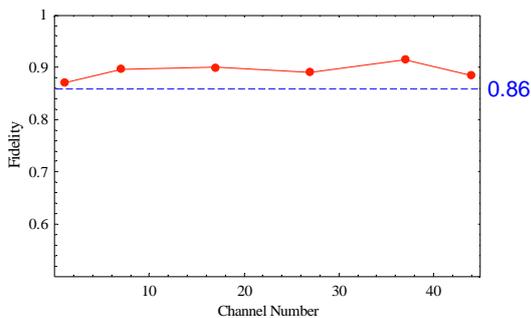

*Figure 3: Experimental result showing fidelity of distributed entangled photons for selected channels.*

Figure 3 shows experimental results. The entanglement fidelities calculated from reconstructed density matrices remain higher than 0.86 for all the selected channels. This shows that the photon-pairs had been successfully distributed. Observed numbers of coincidence counts in 100 seconds (for HH setting) range from 170 at Channel 1 to 250 at Channel 44. Because of randomly varying birefringence of the transmission fiber, we had to use a monitoring laser and polarization controllers to compensate for the polarization drift every few minutes. Polarization stabilization methods such as those demonstrated in [17, 18] should be incorporated in a practical system. We have also found that $\theta$ is wavelength dependent, and so the rotation angle of the HWP must be optimized for every wavelength channel to set $\theta$ to 0. The reason for this characteristic is not understood yet, but we suspect that it is due to polarization-mode dispersion (PMD) of the transmission fiber. Nevertheless, as the compensation can be done after demultiplexing, it does not pose a fundamental problem. It should also be mentioned that the source in our experiment was optimized for one channel and left untouched during all subsequent measurements.

**Conclusion**

In conclusion, we have successfully demonstrated wavelength-multiplexed entanglement distribution over optical fiber using a broadband source of polarization-entangled photon-pairs in the telecom-band. The distribution distance can be extended by replacing the InGaAs single-photon detector modules with super-conducting single-photon detectors [19].

**Acknowledgment**

H. C. Lim was on a postgraduate scholarship awarded by DSO National Laboratories in Singapore.